\documentclass{aa}  
\usepackage{graphicx}

\usepackage{txfonts}
\usepackage{natbib}
\newcommand{\Teff}{T_{\rm{eff}}}
\newcommand{\logg}{\log(g)}
\newcommand{\Halpha}{\rm{H}\alpha}
\begin{document}
  \title{Resolving Stellar Atmospheres I: The $\Halpha$ line and
    comparisons to microlensing observations}

  \author{Christine Thurl \inst{1}, Penny D. Sackett
          \inst{1}
          \and
          Peter H. Hauschildt \inst{2} \fnmsep
          }

   \offprints{C. Thurl}

   \institute{Research School of Astronomy and Astrophysics,
     Australian National University, Mt. Stromlo Observatory, Cotter
     Road, Weston ACT 2611 Australia\\ \email{cthurl@mso.anu.edu.au}
     \and Sternwarte Hamburg, Universit\"at Hamburg, Gojenbergsweg
     112, 21029 Hamburg, Germany } \authorrunning{C. Thurl,
     P.D. Sackett, P.H.Hauschildt} \titlerunning{Resolving Stellar
     Atmospheres} 
\abstract
{We present work on $\Halpha$ spectral line characteristics in
  PHOENIX stellar model atmospheres and their comparison to
  microlensing observations.} {We examine in detail the
  $\Halpha$ equivalent width (EW) and the line shape characteristics
  for effective temperatures of $4500$K$<\Teff < 5600$K where
  $\Halpha$ is a strong spectral feature.} {We find that $\Halpha$ EW
  in models calculated under the assumption of local thermodynamic
  equilibrium (LTE) is up to 15\% smaller than in models without this
  assumption, non-LTE models (NLTE) and that line shapes vary
  significantly for the two model types. A comparison with available
  high quality microlensing data, capable of tracing $\Halpha$
  absorption across the face of one G5III giant, shows that the LTE
  model that fits the EW best is about 100K hotter than and the
  best-fitting NLTE model has a similar $\Teff$ as predicted by the
  spectral type analysis of the observed star but agree within the
  uncertainties of the observationally derived temperature.}{ Neither
  LTE nor NLTE models fit the line shape well. We suspect unmodelled
  chromospheric emission. Line shape diagnostics suggest lower
  gravities than derived for the star and are unacceptable low in the
  case of the LTE models. We show that EW alone is insufficient for
  comparison to stellar model atmospheres, but combined with a new
  shape parameter we define is promising. In stellar parameter ranges
  where the $\Halpha$ line is strong, a NLTE approach of modeling
  stellar atmospheres is not only beneficial but mandatory.}
\date 

   \maketitle
%

\section{Introduction}
Most of our basic knowledge about the interior physics of stars, even
of the Sun, is based on our understanding of stellar atmospheres. The
modeling of such atmospheres allows us to derive the star's
fundamental characteristics such as temperature, gravity, chemical
composition and age. However many outstanding questions still need to
be answered about stellar atmospheres. The ``solar model problem,''
, i.e., the disagreement between model predictions for the neon-oxygen
abundance ratio and measurements by helioseismology, is hotly
debated \citep{drake, martin}. Derivation of elemental abundances
is based on the accuracy of stellar model atmospheres and as such make
and shake our knowledge not only about stellar, but also about
galactic and cosmic evolution \citep{martinreview}.

The main difficulty in constraining stellar atmosphere models with
observations is that most techniques only measure the disk-integrated
flux or disk-integrated spectra of stars. The biggest advance in
one-dimensional (1-D) atmospheres modeling has been to introduce
consistent treatment of Non-Local Thermodynamic Equilibrium (NLTE)
processes \citep{anderson,baron}. However observational tests of NLTE
effects have been difficult to establish, because studying atmospheres
of stars more distant than our Sun has proven difficult and
tedious. 

Various methods have been tried to resolve stars, i.e. measuring their
sizes and resolving surface structure. Direct observations have been
restricted to the Sun \citep[e.g.][]{blackwell} or, when using the
Hubble Space Telescope (HST), to nearby supergiants
\citep[e.g.][]{gilliland}.  Doppler imaging \citep[e.g.][]{olah}
produces observed center-to-limb variations of the integrated flux
that are in good agreement with values derived from current LTE model
atmospheres. Whereas Interferometry \citep[e.g.][]{aufdenberg} show
less limb darkening than predicted by stellar atmospheres models.  Due
to the uneven surface of the moon, surface measurements of nearby
giants via lunar occultations \citep{richichi} have not yet yielded
sufficient precision to distinguish between various model atmospheres.
\citet{odonoghue} and \citet{popper} have used eclipsing binaries to
study surface structure. In such systems both stars are often in a
similar evolutionary state, which complicates tremendously the
decomposition of the single star spectra.  Furthermore, if one star is
evolved and has (almost) filled its Roche Lobe so that mass transfer
occurs, then neither star is a good probe for single-star atmospheres.
However all of these methods are restricted to nearby stars.

Microlensing was proposed to offer an elegant solution to the problem
of observing distant stellar atmospheres \citep{schneider}.  During a
microlensing event a differential magnification pattern across the
face of the source star enables the observer to resolve its surface
(see \S \ref{microlensing} for more detail). Several authors have
studied the effect theoretically.  \citet{heyrov} studied how
microlensing can be used to detect stellar spots on the surfaces of
stars.  \citet{heyrov00} demonstrated the effects of microlensing on
synthesized optical spectra of red giant model atmospheres.
\citet{heyrov03} discussed the difficulty of microlensing light curve
inversion methods.  \citet{bryce} and \citet{hendry} examined the
detectability of star spots via microlensing and their possible
effects on multicolor microlensing light curves.  \citet{bryce03}
examined the change in line profiles due to bulk motion in
circumstellar envelopes during microlensing fold caustic crossing
events.

An early attempt by \citet{lennon96,lennon97} used the
    magnification effect of the microlensing event 96-BLG-3 for
    spectral studies of stellar atmospheres. They further suggested an
    effect of center to limb variations on the $\Halpha$ line profile
    during such an event.
  A number of well-studied microlensing events have resulted in
  measurements that resolve the source star; we discuss one of these
  in more detail in \S \ref{ob02069}. The names and major
  characteristics of these events are listed in Table \ref{events},
  with the respective publications. The major findings
  were promising comparisons of observationally-deduced limb darkening
  coefficients with model atmospheres and variations of equivalent
  width (EW) of the H$\alpha$ line. 
One of these events, EROS BLG-2000-5, is a cusp-crossing event,
    which shows that despite the increased complexity in modeling not
    only fold caustic crossings can be used for these purposes.  For
the most recent event, OGLE-2002-BUL-69 \citep{cassan,kubas}, a
well-sampled light curve and high resolution spectra are available at
crucial times during the event, making it a good target with which to
compare high resolution model atmospheres. Four events of similar
quality have been observed in 2004 and another four in
2005\footnote{http://planet.iap.fr} and many more are to be expected
in the years to come, making microlensing a viable future for testing
stellar model atmospheres.

\begin{table}
\begin{minipage}{\columnwidth}
\label{events}      
\centering          
\begin{tabular*}{\columnwidth}{l l l c c c c c  }     
\hline 
\hline      
Event Name& Publication   & results\\
\hline
MACHO Alert 95-30&\citet{planetma9530}& $\Halpha$ EW \\
MACHO 97-BLG-28&\citet{planetma9728}& limb darkening\\
MACHO 97-BLG-41&\citet{planetma9741}& limb darkening\\
OGLE-1999-BUL-23&\citet{planetog9923}& limb darkening\\
MACHO 98-SMC-1&\citet{planetsmc}& limb darkening\\
EROS BLG-2000-5&\citet{planet}& limb darkening\\
&\citet{castro}&  $\Halpha$ EW \\
&\citet{fields}& \\
MOA 2002 -BLG-23&\citet{planetmo0223}&limb darkening\\
OGLE-2002-BUL-069&\citet{cassan}&$\Halpha$ EW\\
\hline
\end{tabular*}
\end{minipage}
\caption{Observed microlensing events with characteristics and
  results.  In the first column are listed microlensing events
  discussed in greater detail by the authors in the second column.
  The last column lists the main findings for the events,
  i.e. H$\alpha$ equivalent width variations and limb darkening
  coefficients were measured in the respective events.}
\end{table}

This is the first in series of papers aimed at developing methods to
test stellar atmosphere models by confronting them with observations
that resolve stellar surfaces photometrically or spectroscopically.
Based on the most current theoretical models, we describe what
observations can be made in the future in order to distinguish
between various stellar models and learn the most from comparisons to
observations. 

 We lead into this work by describing, in \S\ref{method}, our methods
 and the PHOENIX stellar models. In \S\ref{extract} we explain the
 reasons for and analysis techniques of studying the $\Halpha$
 line. In \S\ref{grid}, we introduce the spectral parameter range
 studied and show that spectral differences between LTE and NLTE
 cannot be ignored within the $\Halpha$ line for stars with $4500K
 <\Teff<5600K$. How microlensing can be used as a tool to resolve the
 surfaces of stars is described in \S\ref{microlensing}, where we
 also present results of our comparison of model atmospheric data to
 microlensing $\Halpha$ EW line data for the event OGLE-BULGE-2002-069.

 In \S\ref{chance}, we illustrate how to most appropriately choose the
 best-fitting model out of a set of models for a given parameter or set of
 parameters. In \S\ref{merit}, we introduce a new line shape
 diagnostic and repeat the analysis similar to the EW analysis. Our
 conclusions are summarized in \S\ref{conclusion}.

\section{Methodology \label{method}}


\subsection{Stellar Models \label{code}}

We study the state-of-art atmospheric models produced by the multi-purpose
code PHOENIX \citep{phoenixI,phoenixII}. These models are static and
assume one-dimensional, spherically-symmetric radiative transfer. We
have chosen PHOENIX because its models can be calculated for all
stellar parameters from dwarfs to giants, for both static and variable
stars. A major advantage of PHOENIX models is that the code can
include self-consistently NLTE effects. LTE assumes that, within a
confined region, absorption and emission of photons occurs at the same
rate and the underlying region's temperature structure follows a black
body. NLTE models relax this assumption so that atomic physics may
produce situations in which local areas need not be in thermal
equilibrium.  Spectra calculated from the PHOENIX models can be
calculated for almost any resolution ranging from the UV to radio
wavelengths.

In order to compare stellar model atmospheres to the observations of
real stars we must ``transform'' the models into observables by
simulating the effects of external phenomena on the
spectra. Microlensing, for example see \S\ref{microlensing}, would be
such a transformation, as it occurs externally to the source star, but
does affect the light collected by the observer as a function of time.
Convolving model data with this transformation produces synthetic
data, which we can then analyze in the same way as we would analyze
observational data. We conduct a comparison between PHOENIX LTE and
NLTE synthetic spectra as well as a comparison to high resolution
observational spectral data from a microlensing event.

\section{H$\alpha$ and its Equivalent Width \label{extract}}
In this work we focus on the effects that can be studied with
observations of the $\Halpha$ line. The $\Halpha$ line is a strong
line for stars with $4500<\Teff < 5600K$ and is therefore a frequent
observational target. Different parts of the $\Halpha$ line are formed
in different regions of a stellar atmosphere, some which are more
affected by NLTE than others \citep{gray}. This makes $\Halpha$ a good
candidate to study possible differences between LTE and NLTE modeling
by testing these differences against observations. Another incentive
to study $\Halpha$ is that published data are available for the
resolved source star during some microlensing events
\citep{planet,cassan}. In this section, we concentrate on the EW of
the line as a figure of merit with which to compare LTE with NLTE
effects, and models with observations.  We adopt the usual definition
of EW:
\begin{equation} 
EW=\int_{line}
1-\frac{F_{\rm{line}}(\lambda)}{F_{\rm{cont}}(\lambda)}d\lambda \label{EWeqn}
\end{equation}
where $F_{\rm{line}}$ is the radiant flux as a function of
wavelength $\lambda$, and $F_{\rm{cont}}$ the value of the flux
from the continuous spectrum outside the spectral line.  The
analytical potential of EW as a tool for studying atmospheric models
depends on one's ability to measure its value consistently for both
model and observational spectra, which requires a careful choice of
the continuum and the rejection of spectral blends.  In order to reduce
systematically-introduced effects to the analysis, we apply the same
algorithm to all our model and observational data. 

\subsection{Extracting Equivalent Width from Model Spectra and Data \label{algo}}

For both model and observational data, the EW is determined as
follows. We first apply a standard vacuum-to-air conversion to all
model data. We then remove blended spectral lines and from this set of
model/data points we determine the continuum around the spectral line
that we wish to analyze and normalize the spectra. Note that by
``continuum,'' we mean the ``pseudo continuum'' determined by fitting,
as there is no way of determining the \emph{real continuum} for
spectral line observations.  After smoothing the data, we obtain the
EW by spline interpolation followed by direct integration.

\begin{figure}
\centering
\includegraphics[angle=-90,width=\columnwidth]{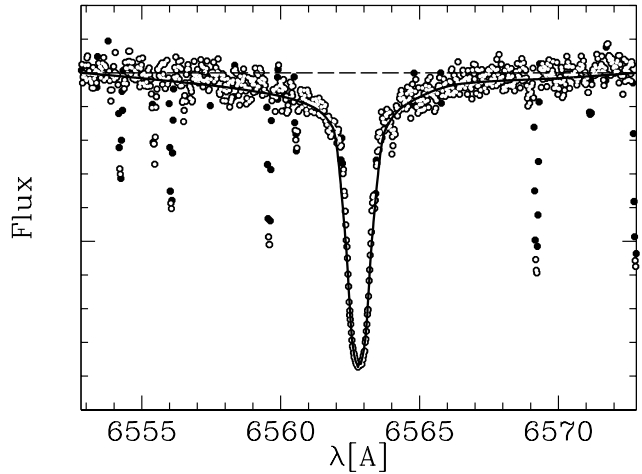}
\caption {Illustration of our EW integration algorithm: black solid
  points are the model output data around the $\Halpha$ line with
  added noise, distributed at random following a Gaussian with a root
  mean squared (rms) scatter $\sigma$ of $2.7 \%$. Marked in white are those
  points chosen by the algorithm to be used for the integration. These
  are then binned, spline interpolated (black solid line), and
  integrated up to the continuum (long dashed line) to give the EW of
  the line.}
 \label{fitting}
\end{figure}

Our clipping algorithm removes lines blended with $\Halpha$ from the
spectral data, which would otherwise contribute to the $\Halpha$ EW
during direct integration. We discard all points from the initial
spectral data (whether model or fully-reduced observational data) that
differ in flux from that of either of their neighbor points by more
than a threshold, $\cal{T}$, which is determined by the
signal-to-noise of the data and its spectral resolution, $\Delta~\lambda$ via, 
%
%
%
\begin{equation}
\cal{T}= \rm{C} \sigma^2/\Delta \lambda \ , 
\end{equation}
where $\sigma$ is the rms scatter of all flux points about their
mean across the wavelength range considered. $C$ is a constant used to
adjust the magnitude of the threshold to ensure optimal clipping.
%
%
We chose $C$ by testing our clipping routine on different
modeled and observed spectra, and found that a different value of C
must be used within the core of the line in order to not reject
spectral points corresponding to the line itself.  After clipping, the continuum is
then determined from the remaining points.

Figure \ref{fitting} illustrates how we obtained the $\Halpha$ EW in
both synthetic and observed data.  We assume the continuum to be
linear over the narrow spectral range considered. We compute the mean
flux value of the outermost 1$\AA$ of the spectra on either side of
the $\Halpha$ line, and fit a continuum line through these
values. Although the synthetic data are virtually noiseless, we
determine the continuum in this way so that the same method can be
applied to the noisier observational data. We have tested and
confirmed this method by determining the EW for integration ranges
from 1$\AA$ to 10$\AA$. The slight increase in EW was linear from
9$\AA$ onwards, indicating that the line signature lies within $\pm$
9$\AA$ and it is safe to determine the continuum outside of this
range.

When smoothing, we divide the spectral data into three
separate sections: (a) $6564.8 \pm 1$$\AA$ to treat the core of the
line, (b) the region excluding the core out to $6564.8 \pm 3$$\AA$, to
cover the wings, and (c) the transition to the continuum of the
spectral range of $6564.8 \pm 10$$\AA$, excluding the core and the
wing regions. Within each of these spectral regions we adapt the sizes
of bins over which we average so as to sample the continuum less
densely than the wings, and the wings less densely than the core.  The
averaged flux values in each bin are then spline interpolated to
give the EW of the line via direct integration.


%

\section{The Stellar Parameter Grid: LTE vs NLTE \label{grid}}

Using this algorithm, we obtained disk-integrated $\Halpha$ EW values
over a grid of 168 PHOENIX LTE and NLTE models.  The stellar
parameters ranged over a grid of $2.0<\logg<5.0$ in steps of $0.5$ and
$4500K < \Teff< 5600K$ in steps of $100$K. A metalicity of
[Fe/H]$=-0.5$ was assumed throughout.
\begin{figure*}
\centering 
\vspace{0.5cm}
 \includegraphics[angle=0,width=16cm]{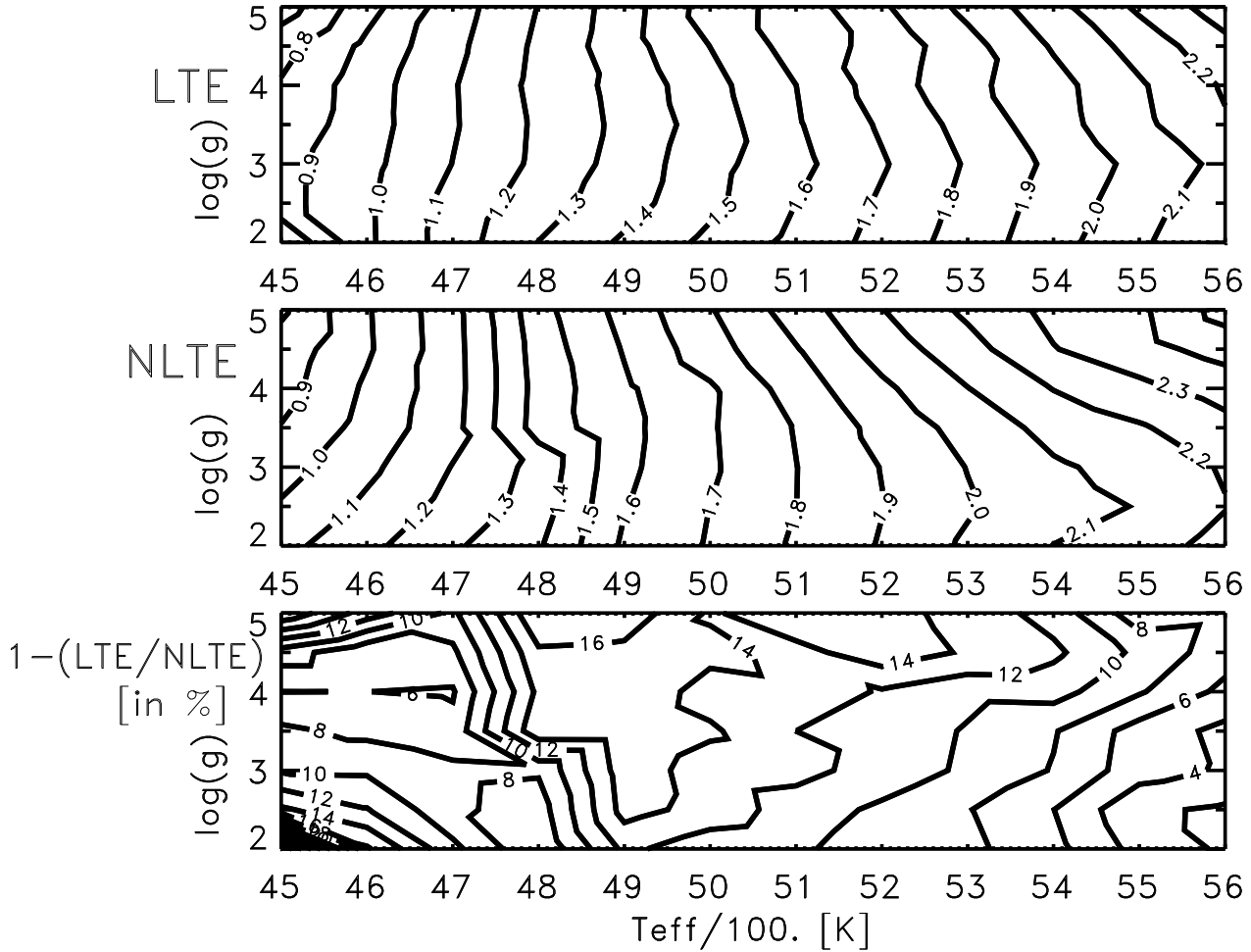}
\vspace{1.5cm}
 \caption{ Upper panel: contours of equal $\Halpha$ EW (integrated
   over the face of the stellar disk) over a range of stellar
   parameters in $\Teff$ and $\logg$ for LTE atmosphere models.
   Middle panel: Same, but for the NLTE atmospheric models. For the
   stellar parameter range shown, NLTE models have higher EW
   compared to LTE models at the same $\Teff$ and $\logg$, as expected
   for the NLTE-affected $\Halpha$ line.  Lower panel: Contours of
   fractional difference $ 1-(EW_{\rm{LTE}}/EW_{\rm{NLTE}})$, given in percent.}
 \label{contours}
  \end{figure*}

In Figure~\ref{contours}, we indicate contour lines of equal
$\Halpha$ EW values. the results for both LTE and NLTE
across the atmosphere model grid.

One can see immediately that $\Halpha$ EW is very sensitive to changes
in $\Teff$, but much less so for changes in gravity.  NLTE models with
higher $\Teff$ show a slightly stronger dependence on gravity than do
the respective LTE models. Previous studies \citep[e.g.][]{gray} have
shown that NLTE calculations produce a deeper core for the $\Halpha$
line than in the LTE regime, which we confirm. The bottom panel shows
the fractional difference in $\Halpha$ EW between the two model types.

Over the middle range of the grid, the fractional difference between
LTE and NLTE is about 15\%.  This shows clearly that, within our
stellar parameter grid, the difference between NLTE and LTE is
significant, and that large systematic errors may compromise any
conclusions drawn for the $\Halpha$ line in this regime if LTE is
assumed for its formation.  This work also indicates that over the
stellar range $\Teff>5400K$ and $\logg<3.0$ a LTE approach may be
reasonable for a study of the $\Halpha$ line, depending on the
precision required.

\section{Microlensing as a Test: Resolving The Surface Of Stars \label{microlensing}}
\begin{figure}
\centering
  \includegraphics[angle=0, width=\columnwidth]{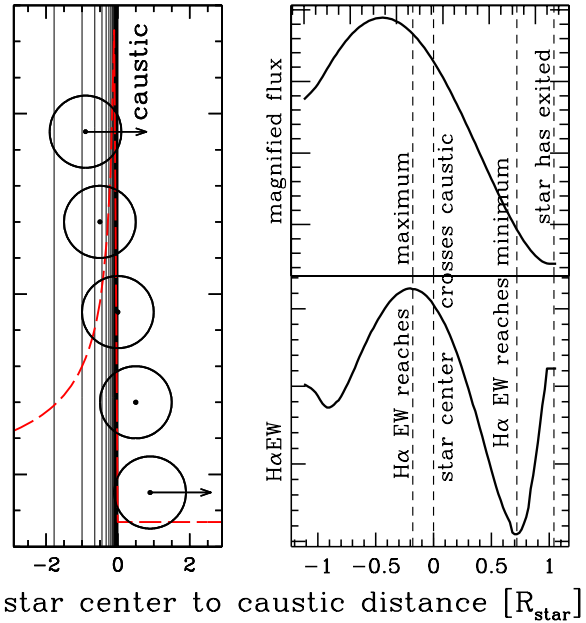}
  \caption{The effect of a microlensing caustic crossing on the
    integrated light and EW of the $\Halpha$ line.: Left panel: As a
    star crosses a caustic, it becomes differentially magnified due to
    the asymmetric magnification pattern (red long dashes). Only the
    exit is shown.  The vertical thin lines show contours of equal
    magnification leading up to the fold caustic line.  Magnification
    drops immediately outside the caustic. Right panels show synthetic
    data for which a stellar model atmosphere has been convolved with
    a microlensing magnification pattern. The top panel shows how the
    total flux received from the star changes due to the magnification of the
    caustic. The bottom panel shows the respective changes in Halpha
    EW for this model. Note the weaker/stronger dips in EW that occur
    when the leading/trailing limb crosses the caustic, contributing
    emission (negative EW) to integrated light.
    \label{lc_ha}}
  \end{figure}

So far we have only compared models of stellar atmospheres.  Can these
differences be observed and, if so, how can this help us to understand
better the underlying physics of line formation and stellar modeling?
In particular, do we have ways to distinguish between NLTE and LTE
effects at the limb and the center of stars? We might expect to
observe larger differences at the limb, where less light originates
from deeper layers of the star, as NLTE processes are more important
in the outer, less dense regions of the atmosphere. Resolving the
stellar surface then becomes crucial.  A fairly new tool of resolving
the surface of stars farther away than the Sun is microlensing
\citep{bryce}. When a stellar object, the lens, crosses the line of
sight to a source star and the lens is a binary, an asymmetric
magnification pattern, which peaks at the so-called caustic, moves
across the face of a source star (see Figure~\ref{lc_ha}). During such
a caustic crossing there is a strongly differential magnification
across the face of the microlensed star, which allows its surface to
be resolved.

Figure \ref{lc_ha} illustrates the microlensing effect on the total
integrated light of the star (upper right panel) and on the $\Halpha$
EW (lower right panel) during a caustic exit.  Light from the parts of
the star directly coincide with the position of the caustic are most
highly magnified, and thus contribute a larger fraction to the
integrated light observed. 
 \begin{figure}
\centering
  \includegraphics[angle=0, width=\columnwidth]{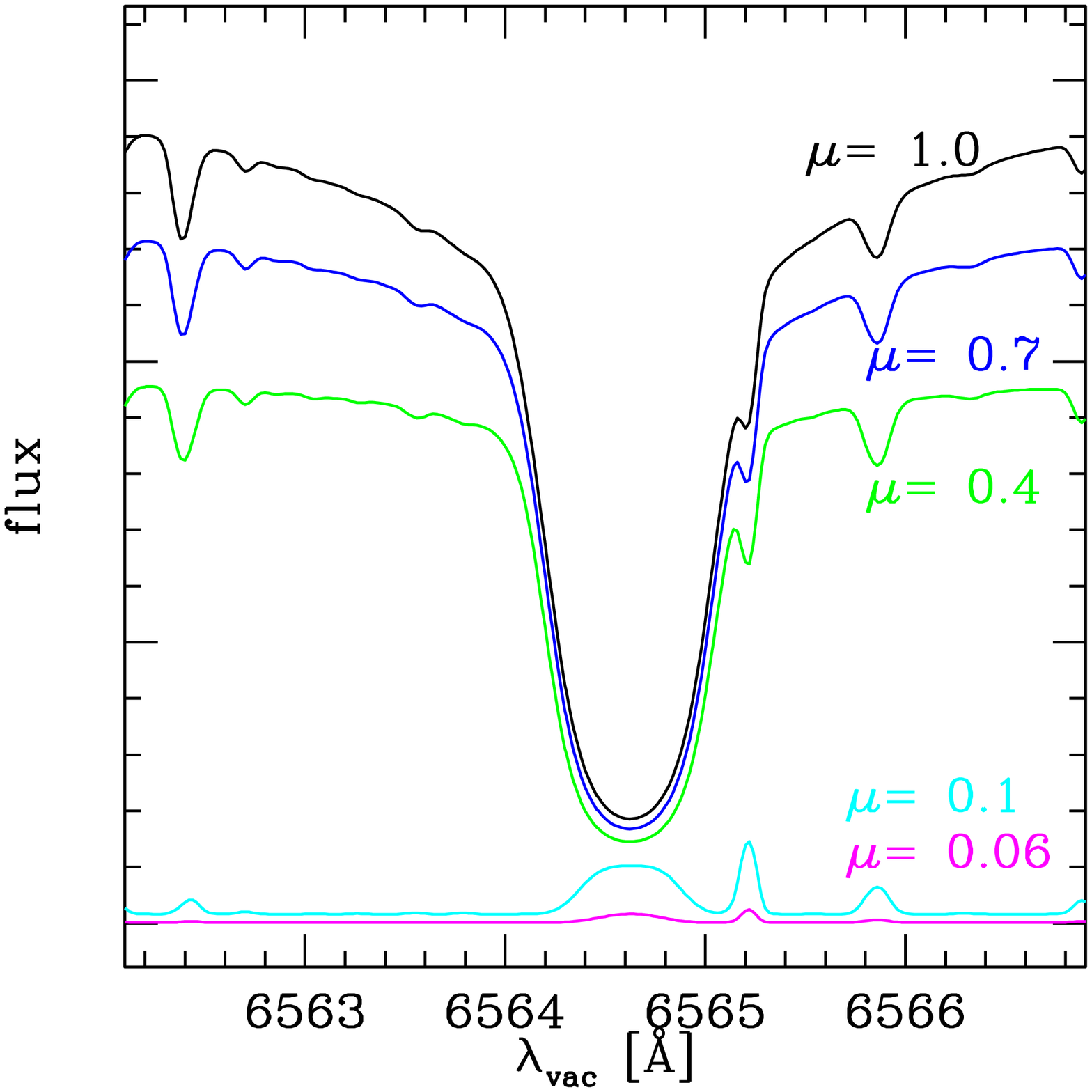}
  \caption{$\Halpha$ spectral lines from stellar model atmospheres for
    different values of $\mu=\cos\theta$, where $\theta$ is opening
    angle to the line-of-sight. At the limb $\mu=0$ and $\mu=1$ in the
    center of the stellar disk. $\Halpha$ is a strong absorption
    feature for most of the stellar disk, but near the limb 
    $\Halpha$ goes into emission. 
\label{halvsmu}}
\end{figure}
Figure~\ref{halvsmu} illustrates the different shapes of the $\Halpha$
spectral line across the face of the star for different values of
$\mu=\cos\theta$, where $\theta$ is opening angle to the
line-of-sight. $\Halpha$ is a strong absorption line in the center of
the star ($\mu=1$) and goes into emission near the limb. This emission
explains the dip in the $\Halpha$ EW during the caustic crossing while
the limb is highly magnified. 

Due to the relative motion of source and lens, the magnification
pattern moves across the source star and produces a light curve of
integrated stellar light, with a typical shape for microlensing
events. The magnification pattern may be rather complex, but in most
cases one can make the assumption that the caustic crossing is a
\emph{fold caustic crossing}, ie., that the caustic has no curvature
across the face of the star. For what follows, we further assume that
the source star is spherically symmetric, and that the relative motion
of source star and caustic is constant and rectilinear. For fold
caustics, a five-parameter fit to the light curve
data is sufficient to extract the magnification pattern of the event
and thus infer the contribution each part of the source star makes to
the integrated light. 
The five parameters describe the time and duration of the event,
    the maximum magnification, how long it takes to reach maximum
    magnification and the underlying linear magnification after the
    caustic crossing \citep{planetsmc}.
Entry into a caustic is unpredictable, but caustic exits can often be
predicted so that proper observations can be put into place. Outside
the caustic, typically there is little or no significant differential
magnification across the stellar disk and the spectrum is similar to
that of the unmagnified star. During the exit, integrated light is
substantially influenced by the highly magnified light from the limb,
only where differential magnification is important.

\subsection{Variations of the $\Halpha$ line EW during a fold caustic crossing: A worked example \label{ob02069}}

Because the $\Halpha$ line is in emission at the limb, the integrated
 EW of the $\Halpha$ line drops during a caustic exit compared to the
 non-differentially magnified case. This occurs first when the leading
 limb crosses the caustic and experiences maximum magnification. The
 effect is stronger when the trailing limb crosses the caustic, as now
 most of the stellar disk has crossed into the low magnification
 region outside of the caustic and contributes a smaller fraction to
 the total integrated light than when most of the star still lies in
 the caustic interior.

During the microlensing event microlensing event OGLE-BULGE-2002-069
the PLANET collaboration \footnote{http://planet.iap.fr} was able to
obtain high resolution VLT UVES data during the caustic exit
\citep{cassan}, and the collaboration kindly provided us with the
fully-reduced spectra around the $\Halpha$ line.

We adopted the parameterization of the fold caustic exit of the event
from \cite{cassan}. For each time at which a UVES spectrum was
available we produced an equivalent synthetic microlensed spectrum for
both LTE and NLTE PHOENIX model atmospheres, for each atmospheric
model on our stellar parameter grid.

We measured the $\Halpha$ EW of the synthetic microlensed data via the
algorithm described in \S\ref{extract}, and applied the same
algorithm to the observational data. Due to the virtual non-existence
of noise in the synthetic data, we adjusted some values of the fitting
parameters for the observational data: the clipping parameter $C$ was
increased, and we have added an extra $8\AA$ outwardly to the range
over which we determine the continuum to achieve a higher stability in
the continuum. Also, since the observed width of the line is not
reproduced by the models, as we describe later, the ``core region'' was
defined as $\Halpha~\pm~0.9$$\AA$ in the data instead of
$\Halpha~\pm~0.75$$\AA$ in the models.

  \begin{figure}
\centering
  \includegraphics[angle=0,width=\columnwidth]{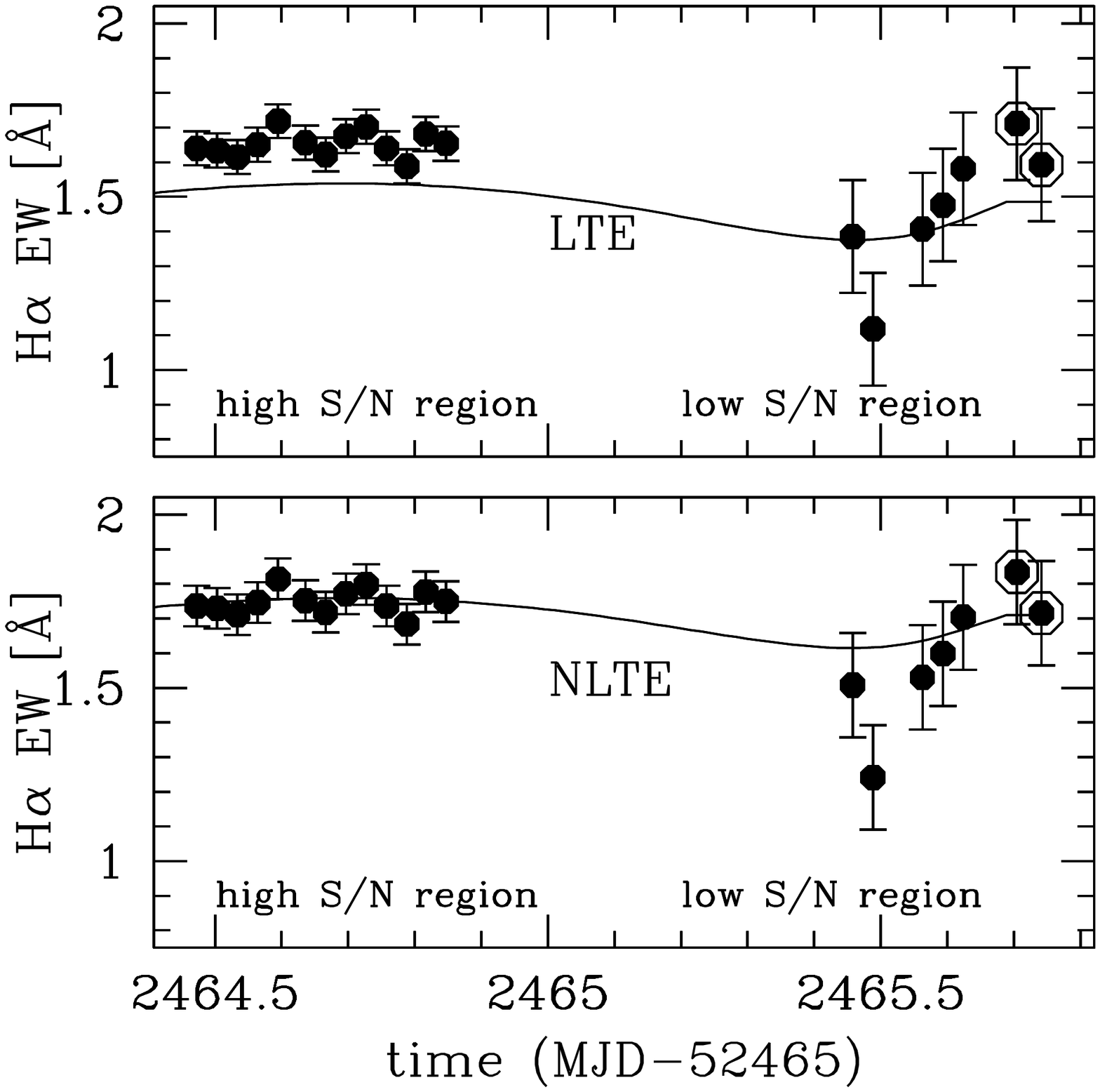}
  \caption{ Changes in the $\Halpha$~EW for OGLE-BULGE-2002-069 during
    the course of its caustic exit.  The solid lines represent the
    synthetic data for LTE (upper panel) and NLTE models (lower
    panel) for $\Teff=5000$K, $\logg = 2.5$ and [Fe/H] = -0.5, taken
    from our model grid, closest to those published by \cite{cassan}.
    The filled circles show our derivation of $\Halpha$~EW from the
    spectra taken by \cite{cassan}. We corrected the observations for
    the systematic uncertainties in the EW determination for LTE and
    NLTE models computed from Monte-Carlo simulations, and applied
    their statistical uncertainties as errorbars.  We did this
    separately for the high and low S/N regions.  The double-ringed
    datapoints show observations taken 36 days after caustic
    exit.} \label{out11}
  \end{figure}
  
Figure \ref{out11} shows our results, where the $\Halpha$~EW is plotted
for each time at which observations have been taken. (The last two
points were taken long after the caustic exit to ensure two baseline points
not affected the differential magnification.)  We plot the $\Halpha$~EW
measurements for only one set of stellar parameters, namely: $\Teff=5000$K,
$\logg = 2.5$ and [Fe/H]$=-0.5$.  These are the parameters from our grid of
stellar models that are closest to the stellar parameters given by
\cite{cassan} $\Teff=5000$K, $\logg = 2.5$ and [Fe/H] $= -0.6$. 
The difference in metallicity of 0.1 dex is well within
    observational errors of the metallicity determination.  Further,
    in previous tests, we did not detect any mesurable changes in
    $\Halpha$~EW from models with [Fe/H] $ = -0.5$ to models with
    solar metallicity, so we are confident in our assumption that
    there are no effects from the 0.1 dex difference in metallicity.

We quantify the uncertainty in our determination of the $\Halpha$~EW,
using a Monte-Carlo approach simulating 999 noisy spectra by adding
random noise to the virtually noiseless synthetic data and rerunning
the algorithm to determine the EW. The simulations have
been restricted to a stellar model of $\Teff = 5000$K and $\logg=
2.5$. We assume no variation of the fractional error distribution across the
whole stellar parameter grid.  The random noise we add follows a
Gaussian distribution with an rms scatter $\sigma$ equivalent to a
signal-to-noise (S/N) of 130 or 50, which \cite{cassan} quote for
their spectra at a resolution of $\lambda/\Delta \lambda = 30000$.
These values translate into a signal-to-noise per spectral data point,
spaced at $0.0174\AA$, of $(S/N)_{\rm{PP}}=36.5$, or $2.7\%$~noise, at
those times where the center of the star is most highly magnified. At
the caustic exit and post-exit time we obtain a $(S/N)_{\rm{PP}} =14$,
or 7\%~noise. When we add this noise to the model spectra at the
appropriate times, the model and observed spectra indeed appear to
have a similar noise level. Thereafter, we shall use \emph{high} and
\emph{low S/N }or \emph{noise} to refer to the 2.7\% and 7\% noise
levels respectively.

From the 999 simulations per noise level and model type, we determined
a histogram of $\Delta\rm{EW}=\rm{EW}_{\rm{model,noisy}}-\rm{EW}_{\rm{model,noise-less}}$from noisy to noise-less data, fitted a
Gaussian to these. The fitting parameters are listed in Table~\ref{errors}.

\begin{table}
\label{errors}      
\centering                          
\begin{tabular}{c c c c c c}        
\hline \hline                 
Simulation&$\mu_{\Delta\rm{EW}} [\AA]$& $\sigma_{\rm{EW}} [\AA]$  &$\sigma_{\rm{EW}}[\%]$ \\    
\hline
   LTE (high S/N) & 0.03379&  0.0492 & 3.28 \\    
   LTE (low S/N) & 0.03175 & 0.1626 &  10.8 \\    
   NLTE (high S/N) & $-0.0623$ & 0.0586& 3.45\\
   NLTE (low S/N) & $-0.0912$ & 0.1509 & 8.88 \\
\hline                                   
\end{tabular}
\caption{Parameters of Gaussian fits to the Monte-Carlo simulation
  results for determining systematic and statistical errors. We have
  separately determined these for LTE and NLTE models at high and low
  noise levels Both models have the same stellar parameters. The
  systematic error, $\mu_{\Delta\rm{EW}}$ is the amount that must be
  subtracted, on average, from the EW determined from noisy ``data''
  compared to that determined from noiseless models. The scatter in EW
  determination about its known mean is $\sigma_{\rm{EW}}$}
\end{table}

The scatter in the observational EW points is similar to the
$\sigma_{\rm{EW}}$ determined from the model spectra.  Large scatter
appears to be mainly a continuum fitting effect. We fit a continuum to
the spectra normalized by \cite{cassan} in the same way as we do it
for the models. Since the observational data is noisy, the continuum
is more uncertain compared to noiseless model data. Even slight
differences in the choice of the continuum will result in a larger
scatter of EW values. Since the data have already been normalized by
\cite{cassan}, for the observations we fit a continuum line with a
fixed slope of zero. Our measurement of the continuum is slightly
lower than that chosen by \cite{cassan}, but is consistent with the
way we pick the continuum for the models.

The EW of the data determined in this way is shown in
Figure~\ref{out11} as dots; in the same figure, model predictions are
also indicated. The overall trend of the observational data is
reproduced by the models, however the offsets between them is obvious.

The LTE models do not match the magnitude of the EW of the $\Halpha$
line of OGLE-BULGE-2002-069 after the caustic exit (double-ringed data
points), where no differential magnification occurs and also at times
when the center of the source star crosses the caustic. This could indicate
that either the stellar atmosphere models are insufficient
and/or we have chosen the wrong stellar parameters from the grid.

\section{Giving the models a fair chance \label{chance}}

Before condemning the model atmospheres prematurely, we need to be
sure that we have chosen models that best fit the $\Halpha$ EW data within 
uncertainties in the determination of $\Teff$ and $\logg$. We quantify
the discrepancy between models and observations by defining $\chi^2$,
the goodness-of-fit figure of merit. 
\begin{equation} \chi^2= \sum_{t_{\rm{obs}}}
  \left[EW_{\rm{obs}}(t_{\rm{obs}})-EW_{\rm{model}}(t_{\rm{obs}})
  \right]^2/(\sigma_{\rm{EW}}^2(t_{\rm{obs}}) \ ,
\end{equation} 
summing over all points in time $t_{\rm{obs}}$ at which an observed
spectrum has been taken. We assume that any systematic errors
introduced by our fitting algorithm and statistical uncertainties in
fitting model spectra are the same across the model grid. Thus, for
$\sigma_{\rm{EW}}$ we use the values derived in \S\ref{ob02069},
namely, around 10\% and 3.5\% for the high and low noise data,
respectively.

\begin{figure*}
\centering
  \includegraphics[angle=90,width=12cm]{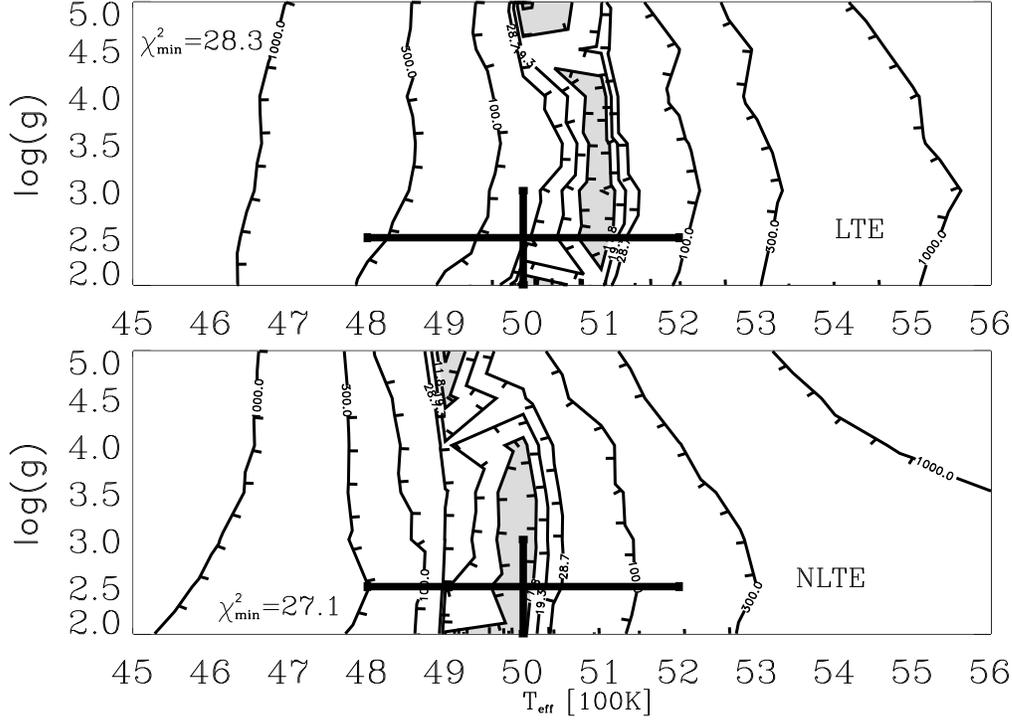}
\vspace{1.cm}
  \caption{ Upper panel: Contours of equal $\Delta\chi^2= \chi^2 -
    \chi^2_{\rm{min}}$, where $\Delta\chi_{\rm{min}}^2$ is minimal
    when best agreement between LTE models and observations of the
    $\Halpha$ EW of OGLE-BULGE-2002-069 are achieved. Lower panel:
    Same for NLTE.  $\Delta\chi_{\rm{min}}^2$ is different for LTE and
    NLTE models and its value is shown in the left corner of the
    figure panels.  The grey area indicates all
    models that lie within $3\sigma$ of the best fitting
    model for two degrees of freedom, namely $\Teff$ and
    $\logg$. Further contour lines are the $4\sigma$, $5\sigma$ and higher
    boundaries. The tick marks on the contour lines indicate the
    direction of decrease in $\Delta\chi^2$.  From their spectral
    analysis, \cite{cassan} determined the stellar parameters to be
    $\Teff=5000$K, $\logg=2.0$, [Fe/H]=-0.6. We mark this choice of
    parameters with the large cross with uncertainties of $\Delta\Teff
    = \pm 200$K and $\logg=\pm 0.5$ as given in \citet{minniti} for
    this type analysis.}
\label{chiTES} 
\end{figure*}

We calculate $\chi^2$ for each model in our model stellar grid, and
show in Figure \ref{chiTES}, contours of equal $\Delta\chi^2$,
\begin{equation}
\Delta\chi^2= \chi^2 - \chi^2_{\rm{min}} \ ,
\end{equation}
where $\chi^2_{\rm{min}}$ is the minimal value of $\chi^2$ across the
model grid, i.e. the goodness-of-fit for the best fitting
model. Models along a contour line fit the observed measurements
equally well. In Figure~\ref{chiTES}, we superimpose on the model grid the
stellar parameters determined by \cite{cassan} for
OGLE-BULGE-2002-069.  We follow an analysis of \citet{minniti}, used
by \citet{cassan}, indicating uncertainties in determining stellar
parameters of $\pm200$K in $\Teff$, $\pm 0.5$ in $\logg$, and $\pm 0.3$
in [Fe/H].

Within the uncertainties of the stellar model, LTE models with
$\approx 100K$ higher $\Teff$ than and NLTE models at the same values
quoted by \citet{cassan} provide the smallest $\chi^2$. Gravity is
not well constrained. The overall goodness of fit is similarly poor
for both LTE and NLTE models at a values of $\approx 28$ for 20
observations, which could be caused by random noise, with only about 7\%
confidence. Only the best fitting models should be examined to study
further differences to the observations. Note that the best-fitting
LTE and NLTE models have {\emph{different temperatures}}.

\section{The Importance of Line Shape \label{merit}}
\begin{figure*}
\centering
  \includegraphics[angle=0,width=\textwidth]{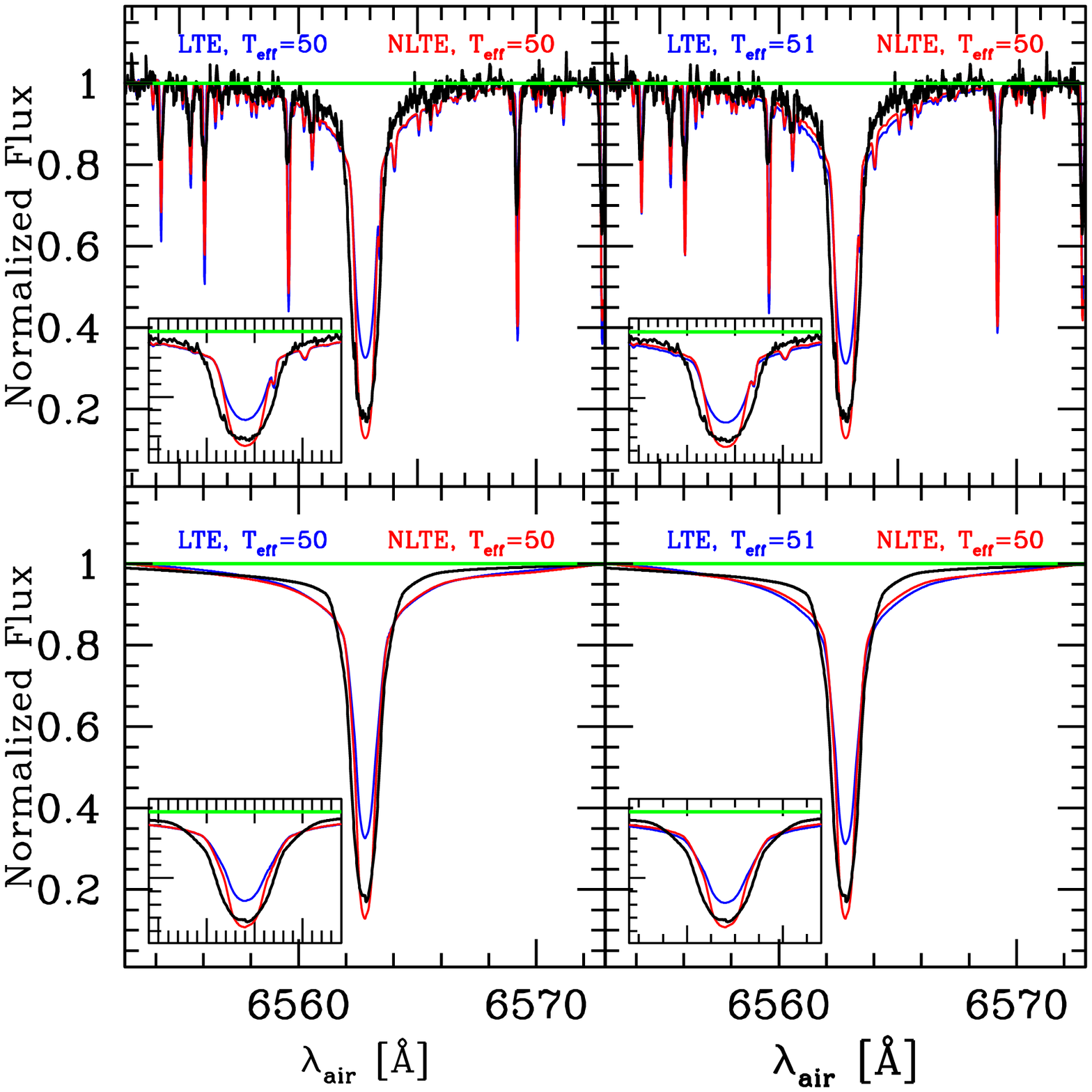}
  \caption{Differences in the spectra between observations of the
    $\Halpha$ line for OGLE-BULGE-2002-069 (in black) and models (LTE
    in blue, NLTE in red) at the same magnification. The left panel
    shows comparison to the models with the stellar parameters closest
    to the published values, the right panel shows the spectra from
    models as chosen from Figure~\ref{chiTES}.  Bottom panels present
    the clipped and spline interpolated lines after the observations
    and models have passed through the EW algorithm from
    \S\ref{algo}. Each inset shows a narrower region around the core
    of the line. For all diagrams $\logg=2.5$ and $\Teff$ is given per 100K.
\label{halpha}}  
\end{figure*}

Equivalent width collapses all the information about a spectral line
into one value that only considers the area of the line under the
continuum, but is independent of the line shape. This means that, 
despite a good agreement between the EW of a model spectral line, and
of the observed spectral line any variations in the line shape will be
ignored if this is the only figure of merit used to test models.

Examples of the $\Halpha$ spectral line (top panel) and their fit from
the algorithm (bottom) panel are shown in Figure~\ref{halpha} for one
observation. The line from corresponding differentially magnified LTE
and NLTE models with parameters as chosen from \cite{cassan} are shown
in the left panels, right panels give the best fitting stellar models
as determined in Figure~\ref{chiTES}. Although the changes appear
small, allowing stellar parameters $\Teff$ and $\logg$ to vary within
the observationally allowed uncertainties, can alter the EW by
10-15\%, as can been seen in Figure~\ref{contours}.

That the models do not generally fit the width of the $\Halpha$ line
(see Figure~\ref{halpha}) suggests that there are other physical
processes occurring which are yet to be explained by stellar
models. One explanation may be chromospheric influences, which
\cite{cassan} claim to have detected in their data. Since $\Halpha$ is
a chromospheric line but the standard model atmospheres do not include
chromospheric calculations, more research is needed to explain this
phenomenon.

This comparison of individual spectra clearly shows where discrepancies in EW
occur, primarily in the wings of the line, but for LTE models also in the
core of the line. However, the differences in shape in the core and the
wings balance each other out when only total EW is considered. The
discrepancies due to the different line shapes are washed
out. Therefore we have devised a different diagnostic to compare model
spectra with observational data by defining the line shape diagnostic,
$\psi^2$,
\begin{equation}
\psi^2(t_{\rm{obs}})=\int_{\rm{line}}\left[{\cal{F}}_{\rm{obs,o}}(\lambda,t_{\rm{obs}})-{\cal{F}}_{\rm{model,o}}(\lambda,t_{\rm{obs}})\right]^2
  d\lambda \ ,
\end{equation}
where we integrate the square of the flux difference that results when
we subtract the model spectrum from the observational spectrum, after
both have been normalized, clipped and spline-interpolated. Note, that
the units of this new diagnostic are $\AA$, since the flux has already
been normalized. Similarly, we can obtain ${\psi^2}$ for line
shape comparisons between LTE and NLTE models. 
\begin{figure}
\hspace{-0.8cm}
  \includegraphics[angle=0,width=\columnwidth]{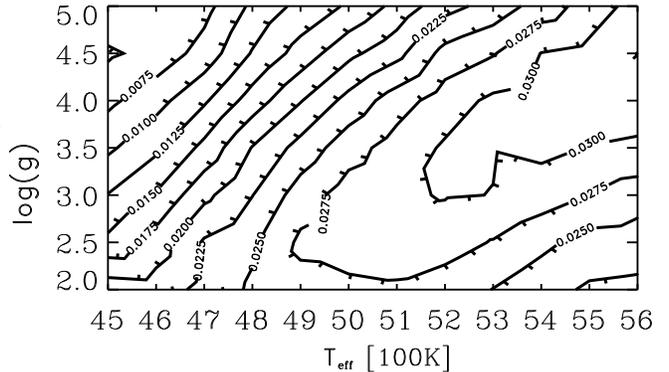}\\
\vspace{-1.8cm}
  \caption{ Contours of equal ${\psi^2}$ for a NLTE-LTE model
    comparison.  The tick marks on the contour lines indicate the
    direction of decrease. Measured by this quantity, line shapes for
    NLTE and LTE models differ least for lower $\Teff$ and higher
    $\logg$ within the parameter grid. Larger differences are found
    for higher $\Teff$ and higher $\logg$.}
\label{shape}  
\end{figure}
The result of which is shown in Figure~\ref{shape} for non-microlensed
model data. In order to compare LTE and NLTE models to observational
data taken during a microlensing event, and to find the best-fit model
based on the measured shape parameter, we define a new shape figure of
merit
\begin{equation}
\Psi\equiv\sum_{t_{\rm{obs}}}{\psi^2}(t_{\rm{obs}}) \,
\end{equation}
 which is sum of equally weighted ${\psi^2}$ over each observation
 during the event. The units of $\Psi$ are still $\AA$.

\begin{figure*}
\centering \includegraphics[angle=90,width=12cm]{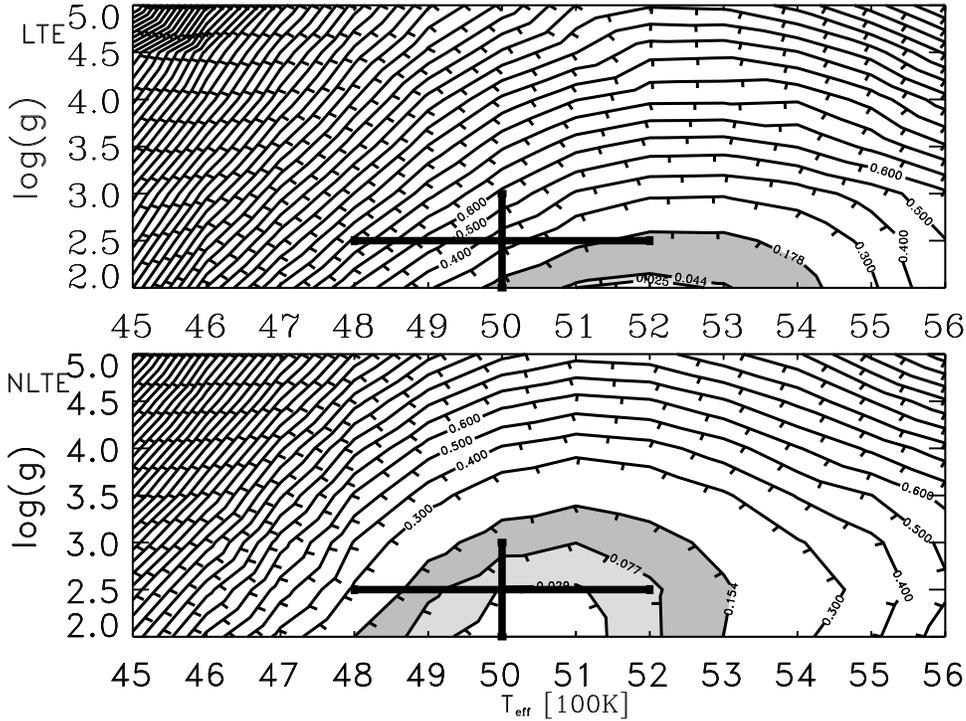}\\
\vspace{1.0cm}
  \caption{Contours of equal $\Psi -\Psi_{\rm{min}}$ for comparison of
    the observational data with LTE models (upper panel) and NLTE
    models (lower panel). The contours clearly constrain not only the
    $\Teff$ but also the $\logg$ of the best fitting model parameters.
    This means that the best fitting LTE model has not only a higher
    $\Teff$ and a lower $\logg$, whereas the best fitting NLTE model
    has stellar parameter closest to the one suggested by
    \citet{cassan}, indicated by the black cross.
\label{shapeint}} 
\end{figure*}

Defining the shape parameter $\Psi$ allows us to collapse the
available time series data to one quantity per stellar model, while
conserving the information of the line shape comparison. In effect,
$\Psi$ is sum of $\chi^2$ comparison of all the spectra themselves,
across a wavelength range defined by the line, equally weighted.  We
determine uncertainties in $\Psi$ by fitting a Gaussian to the
distribution of $\Psi$ values resulting from 999 Monte-Carlo
simulations. In each simulation we sum up 13 low noise and 7 high
noise $\psi^2$ values randomly drawn from the distribution of $\psi^2$
values, which we have determined via 999 Monte-Carlo simulations in a
similar fashion as for EW in \S\ref{ob02069} for low and high noise,
LTE and NLTE models. We fit the resulting distributions of $\psi^2$
with a combination of a Gaussian and a straight line as the
distribution is asymmetric because $\psi^2$ is positive definite. We
fit the $\Psi$ distribution with a Gaussian convolved with a straight
line. From this we deduce 3, 2 and
1$\sigma$ levels of confidence which we mark in Figure~\ref{shapeint}
as dark grey, light grey and the enclosed white areas.  

In Figure~\ref{shapeint}, we show contour plots of equal $\Psi-
\Psi_{\rm{min}}$, where $\Psi_{\rm{min}}$ is the lowest value of
$\Psi$ across the model grid.  For LTE models this is
$\Psi_{\rm{min}}=0.71$, for NLTE models $\Psi_{\rm{min}}=0.47$. The
shape parameter ${\psi^2}$ is much more dependent on $\logg$ than is
EW, which was strongly dependent only on $\Teff$.
The dark grey shaded area indicates the 99.73\% level of confidence
around the best fitting model. The enclosed light grey and white areas
show the 95.4\% and 68.3\% confidence levels, respectively. The
best-fitting LTE model has a $\logg$ of less than 2.0, substantially
less than what the observationally-determined stellar parameters
suggest. As with fits to EW, LTE fits to the shape parameter, $\Psi$,
again suggest a higher $\Teff$ by about 200K, i.e. 100K higher than
the EW-best-fitting LTE model suggested.  Although the best-fitting
NLTE model for the shape parameter leans towards lower gravities,
these are in agreement with the observationally determined stellar
parameters. The overall fit to $\Psi$, however, is fairly poor: The
distribution of $\Psi$ at random shows that $99.9\%$ of values drawn
randomly from the distribution of $\psi^2$ for NLTE models suggests a
value below 0.07.  For LTE models this reduces to a value less than
0.01.

\section{Conclusions \label{conclusion}}
As a part of a larger program to compare model atmospheres to
observational data, we have developed an algorithm to consistently and
automatically determine and compare spectral line characteristics.
Here, we have focused on equivalent width (EW) of the $\Halpha$ line
and also introduced a line shape parameter $\psi^2$, which
emphasizes flux differences between two spectra. We have studied a
stellar parameter range of $4500<\Teff<5300$ and $2.0<\logg<5.0$ at
[Fe/H]$=-0.5$, a range typical of microlensed source stars in the
Galactic Bulge.

We have determined the uncertainty of this algorithm by performing
Monte-Carlo simulations in which noise is introduced to a model
spectrum and the extraction procedure repeated. The simulations have
been restricted to a stellar model of $\Teff = 5000$K and $\logg=
2.5$. We assume no variation of the fractional error distribution across the
whole stellar parameter grid. The EW of the $\Halpha$ spectral line as
predicted by PHOENIX atmospheric models is substantially different for
LTE and NLTE model. The wrong choice of model type may result in
uncertainties in EW larger than 8\% for most models with
$4500<\Teff<5300$ and $2.0<\logg<5.0$.  The shape parameter $\psi^2$
in an LTE-NLTE comparison is statistically significant and largest for
high $\Teff$ and high $\logg$.

Choosing the right model to compare to the observations is crucial
when one wants to understand stellar atmosphere modeling.  As an example,
we study $\Halpha$ spectral data from the microlensing event
OGLE-BULGE-2002-069. First, we determine the best-fitting model to the
$\Halpha$ EW data over the range of our stellar grid for both LTE and
NLTE models over the course of the caustic exit, during which the
source star is resolved. We find that in order to fit the observed
data an LTE model requires a $\Teff$ about 100K larger than that
proposed by \citet{cassan} based on spectral type analysis for this
source star. The best-fitting NLTE model, on the other hand, requires
a $\Teff$ as proposed by \citet{cassan}.  The
absolute $\chi^2$ is lower for NLTE PHOENIX models than for LTE
models, suggesting a slight preference for the NLTE model.

Second, we determine the best-fitting model to the $\Halpha$
$\psi^2$ which we define, and find that the best-fitting LTE
model has a $\Teff$, that is 200K higher than the observationally
determined $\Teff$ and a significantly lower $\logg$, The best-fitting
NLTE model agrees with the observationally determined values for
$\Teff$ and $\logg$ within the uncertainties, although it tends to
underestimate both $\Teff$ and $\logg$ for this star.  The NLTE models
produced a better figure of merit than the LTE models for both diagnostics.

We caution, however, that declaring either LTE or NLTE models a
failure based on analysis of $\Halpha$ only is premature in the case
of OGLE-BULGE-2002-069 .  Chromospheric effects may alter the
characteristics of the $\Halpha$ line significantly, and the
atmosphere models used here do not include chromospheres. Indeed,
\citet{cassan} claim to have detected an anomaly in the $\Halpha$ line
at some times during the caustic crossing that may be due to a
chromosphere.

We would like to point out that with our method we are already placing
strong constraints on $\Teff$ and $\logg$ by using only two parameters
measured for one particular line. Future studies of multiple lines and
multiple diagnostics will provide even stronger constraints. In
subsequent work, we will determine the most effective choices of
spectral lines and their diagnostics for testing model atmospheres and
include chromospheric calculations in the models for direct comparison
to observations of OGLE-BULGE-2002-069.

\begin{acknowledgements}
We would like to thank John Tonry for kindly sharing with us his
Marquardt-fitting and minimization routine. We would also like to
thank the members of PLANET for providing us with the valuable UVES
spectra for OGLE-BULGE-2002-069 and fruitful discussions of the
matter.  This work was supported by ANU Supercomputing facility
(APAC). We would like to thank the National Facility help desk, and in
particular Ben Evans, for support and help with the APAC system. We are
indebted to Martin Asplund for helpful discussions in preparation of
this paper.
\end{acknowledgements}

\end{document}